\documentclass[twocolumn,%showpacs,
prl
,superscriptaddress
]{revtex4}

\usepackage{graphicx}
\usepackage{amssymb}
\usepackage{amsmath}
\usepackage{color}
\newcommand{\ve}[1]{{\mathbf #1}}
\def\lba{\left(}    \def\rba{\right)}
\def\lbc{\left[}    \def\rbc{\right]}

\begin{document}

%\linenumbers

\title{Density instabilities in a two-dimensional dipolar Fermi gas}

\author{M. M. Parish}
%\email{mmp24@cam.ac.uk} %
\affiliation{Cavendish Laboratory, JJ Thomson Avenue, Cambridge,
 CB3 0HE, United Kingdom} %
\affiliation{London Centre for Nanotechnology, Gordon Street, London, WC1H
0AH, United Kingdom}

\author{F. M. Marchetti}
%\email{francesca.marchetti@uam.es} %
\affiliation{Departamento de F\'isica Te\'orica de la Materia
Condensada, Universidad Aut\'onoma de Madrid, Madrid 28049, Spain}

\date{\today}

\begin{abstract}
  We study the density instabilities of a two-dimensional gas of
  dipolar fermions with aligned dipole moments. The Random Phase
  Approximation (RPA) for the density-density response function is
  never accurate for the dipolar gas, and so we incorporate
  correlations beyond RPA via an improved version of the
  Singwi-Tosi-Land-Sj\"{o}lander scheme. In addition to density-wave
  instabilities, our formalism captures the collapse instability that
  is expected from Hartree-Fock calculations but is absent from
  RPA. Crucially, we find that when the dipoles are perpendicular to
  the layer, the system spontaneously breaks rotational symmetry and
  forms a stripe phase, in defiance of conventional wisdom.
\end{abstract}


\maketitle

%\paragraph{Introduction ---}
Ultracold atomic gases have thus far provided a veritable playground
in which to explore quantum many-body phenomena.
One of the field's great successes is the ability to tune the
effective interatomic interactions via Feshbach resonances, thus
allowing one to access the regime of strong correlations in a
controllable manner.
Furthermore, the ability to create tightly bound heteronuclear
Feshbach molecules with a permanent electric dipole moment provides a
promising system in which to study many-body physics with long-ranged
dipole-dipole interactions (for a review see, e.g.,
Ref.~\cite{Baranov2008}). Indeed, such polar molecules can have
interactions that are several orders of magnitude larger than those
for atomic magnetic dipoles~\cite{lahaye2009}.

Of particular interest are fermionic polar molecules confined in
two-dimensional (2D) geometries: fermionic ${}^{40}$K${}^{87}$Rb
molecules have been very recently created~\cite{Ni2008}, cooled down
to quantum degeneracy~\cite{Ni2010}, and their lifetime increased by
the confinement in 2D~\cite{miranda2011}.
However, correlations are expected to be enhanced in 2D compared to
3D, and thus a major challenge is how to describe theoretically such
correlations in the dipolar system.  It is this issue that we will
address in this Letter.

\begin{figure}
\centering
\includegraphics[width=0.9\linewidth,angle=0]{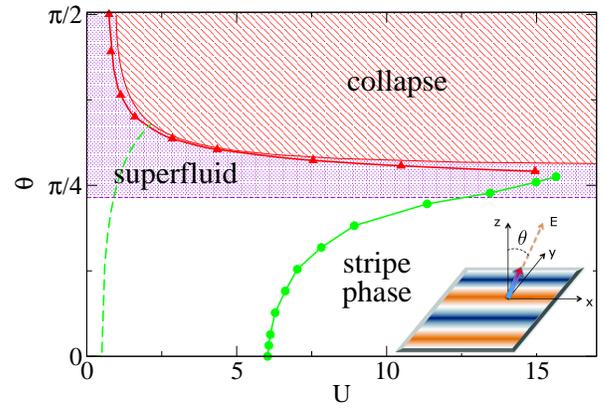}
\caption{(Color online) Phase diagram for a 2D dipolar Fermi gas as a
  function of the dipole orientation angle $\theta$, as defined in the
  inset, and dimensionless interaction strength $U = mD^2
  k_F/\hbar^2$, where $D$ is the dipole moment and $k_F$ is the Fermi
  wave vector. The (green) circles mark the transition to a stripe
  phase while the (red) triangles correspond to the collapse
  instability, all derived using our improved STLS formalism. For
  comparison, we include the RPA result for the stripe instability
  (dashed [green] line), the Hartree-Fock result for the collapse
  instability (stripe pattern [red] region), and the area where
  $p$-wave superfluidity is expected to occur (shaded [violet]
  region), as predicted by Ref.~\cite{bruun2008}.}
\label{fig:phase_diag}
\end{figure}

We focus on a dipolar Fermi gas in a \emph{single} layer,
where the dipole moments are all aligned by an external electric field
$\ve{E}$, making an angle $\theta$ with respect to the normal of the
2D $x-y$ plane (inset of Fig.~\ref{fig:phase_diag}).
Even for this simple 2D geometry, the anisotropic interactions provide
an exotic twist to the problem and a rich phase diagram is
expected. For sufficiently large tilting angles, $\theta> \arcsin
(1/\sqrt{3})$, the interaction develops an attractive sliver in the
plane, eventually leading to $p$-wave
superfluidity~\cite{bruun2008}. For small tilting angles $\theta \neq
0$, the repulsive, anisotropic interaction
% its anisotropy provides an exotic twist to the problem, 
is expected to give rise to anisotropic density-wave (stripe)
phases~\cite{DasSarma2010,yamaguchi2010}, before eventually yielding a
Wigner crystal at sufficiently high densities and/or strong
interactions~\cite{Cremon2010}.
However, in Refs.~\cite{DasSarma2010,yamaguchi2010}, the basic
description of the stripe phase is derived from the Random Phase
Approximation (RPA) for the density-density response function, and
this is not expected to be accurate for the 2D dipolar Fermi gas: As
well as neglecting the exchange correlations resulting from Fermi
statistics,
%which is known to give an incorrect short range behaviour, in addition, 
RPA fails to correctly describe the long wavelength regime of the
density-density response function, unlike in the case of the 2D
electron gas. Furthermore,
RPA does not settle the question of whether or not the 2D dipolar
Fermi gas spontaneously breaks rotational symmetry and forms a stripe
phase for \emph{isotropic} interactions ($\theta =0$), which is of
fundamental interest to other quasi-2D systems such as the cuprate
superconductors~\cite{kivelson1998}.

In this Letter, we include correlations beyond RPA using an improved
version of the Singwi-Tosi-Land-Sj\"{o}lander (STLS)
scheme~\cite{STLSpaper}, which has had much success in describing
electron systems~\cite{vignale_book}.
Using this formalism, the effect of correlations is evident in the
pair correlation function, where we observe a ``correlation hole''
forming around each fermion with increasing interaction. We map out
the instabilities of the density-density response function and we see
the existence of a stripe phase, similarly to RPA, though for
considerably larger dipole strengths and/or densities. However, in
contrast to RPA, we also observe a collapse instability for
sufficiently large $\theta$, which is consistent with Hartree-Fock
calculations~\cite{bruun2008,yamaguchi2010,babadi2011,sieberer2011}. Last
but not least, we show that the system does indeed spontaneously break
rotational symmetry to form a stripe phase when $\theta=0$.

%\paragraph{Methods ---}
%dipolar potential
The effective 2D dipolar interaction for aligned dipoles confined in a
layer of width $W$ can be evaluated as per
Ref.~\cite{fischer_06}. Parametrizing the $x$-$y$ in-plane
  momentum in polar coordinates $\ve{q}=(q,\phi)$ ($\phi = 0$
  corresponds to the direction $x$ of the dipole tilt in the inset of
  Fig.~\ref{fig:phase_diag}), in the limit $q W\ll 1$ (the expected
regime of the experiments), the 2D interaction can be written as:
\begin{equation}
  v (q,\phi) = V_0 - 2\pi D^2 q\left(\cos^2\theta - \sin^2\theta
  \cos^2\phi\right)\; ,
\label{eq:dipolar_int}
\end{equation}
where $D$ is the dipole moment.
$V_0$ corresponds to the short-ranged contact interaction, which
depends on the confinement,
%$V_0 \propto D^2/W$. 
and the confinement width $W$ provides a natural cut-off for the
quasi-2D system: $\Lambda \sim 1/W \gg k_F$.
The dipolar system is parametrized by the angle $\theta$ and the
dimensionless interaction strength $U = mD^2 k_F/\hbar^2$, where $m$
is the fermion mass and $k_F = \sqrt{4\pi n}$ is the Fermi wave vector
($n$ is the density). Note that the effective coupling increases with
increasing density, in contrast to the case of Coulomb interactions,
where the regime of strong coupling corresponds to low densities.

%Linear response theory
In the following, we analyse the inhomogeneous phases of a 2D dipolar
Fermi gas using the linear response theory. Here, the linear density
response $\delta n(\ve{q},\omega)$ to an external perturbing potential
$V^{ext}(\ve{q},\omega)$ defines the density-density response function
$\chi(\ve{q},\omega)$ in frequency and momentum space:
\begin{align}
  \delta n(\ve{q},\omega) = \chi(\ve{q},\omega) V^{ext}(\ve{q},\omega)
  \; .
\label{eq:linear_response}
\end{align}
In the static limit, $\omega = 0$, the appearance of a divergence in
$\chi$ at a particular wave vector $\ve{q}_c$ signals an instability
towards the formation of a density wave with period set by $\ve{q}_c$.
Note that if the instability only depends on the magnitude $q_c \equiv
|\ve{q}_c|$ and is insensitive to the angle $\phi$, then the
inhomogeneous phase may consist of multiple density waves, so that it
forms, e.g., a triangular lattice rather than a stripe phase.

In addition to density instabilities, we can use the
fluctuation-dissipation theorem to extract the ground-state
correlation functions from $\chi(\ve{q},\omega)$.  A standard quantity
is the pair correlation function
%for a uniform system is defined as
$g(\ve{r_2}-\ve{r_1}) = \frac{1}{n^2}\langle
\psi^\dag(\ve{r_2})\psi^\dag(\ve{r_1}) \psi(\ve{r_1})
\psi(\ve{r_2})\rangle$, where $\langle \cdots\rangle$ is the
expectation over the ground state and $\psi^\dag(\ve{r})$ is the
creation operator for a spinless fermion at position $\ve{r}$.
%Indeed, $g(\ve{r})$ uniquely determines the interaction energy for
%the interacting system.
This is related to the static structure factor $S(\ve{q})$ by:
\begin{align}
  g(\ve{r}) = 1 + \frac{1}{n} \int \frac{d\ve{q}}{(2\pi)^2}
  e^{i\ve{q}.\ve{r}} \lbc S(\ve{q})-1 \rbc \; .
\label{eq:pair_corr}
\end{align}
which, in turn, is connected to the response function via
\begin{equation}
  S(\ve{q}) = -\frac{\hbar}{n\pi} \int^{\infty}_{0} d\omega
  \chi(\ve{q},i\omega) \; .
\label{eq:struct_fact}
\end{equation}
Note that here the integration is performed along the imaginary
frequency axis.

\begin{figure}%[htbp]
\centering
\includegraphics[width=0.8\linewidth,angle=0]{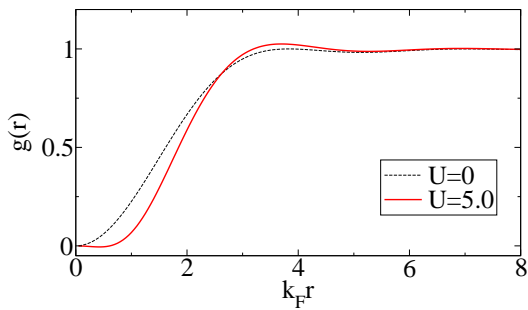} %
\caption{(Color online) Pair correlation function $g(r)$ of a 2D
  dipolar Fermi gas at $\theta = 0$ with and without
  interactions. Note that we must always have $g(0)=0$ for a gas of
  identical fermions. Increasing the repulsive interaction $U$
  decreases the likelihood of two fermions being close together, thus
  resulting in a ``correlation hole'' near $r=0$ for $U>1$.}
\label{fig:pair_corr}
\end{figure}

For a non-interacting 2D Fermi gas at zero temperature, the response
function can be evaluated exactly~\cite{stern67},
\begin{equation*}
  \Pi(\ve{q},i\omega) = \frac{m}{2\pi b} \left\{\sqrt{2} \left[a +
    \sqrt{a^2 + \left(\omega b\right)^2} \right]^{1/2} -b\right\}\; ,
\end{equation*}
with $a=\frac{b^2}{4} - b\frac{k_F^2}{m} - \omega^2$ and
$b=\frac{q^2}{m}$.
If we insert $\Pi(\ve{q}, i\omega)$ into \eqref{eq:struct_fact} to
obtain the non-interacting structure factor $S_{0}(\ve{q})$, and then
use Eq.~\eqref{eq:pair_corr}, we find that $g(0) = 0$ (see
Fig.~\ref{fig:pair_corr}), as expected from Pauli exclusion.

%Introduce RPA for a general interaction $v(\ve{q})$
For fermions interacting via a two-body potential $v(\ve{q})$, one
often relies~\cite{DasSarma2010,yamaguchi2010} upon the RPA to
estimate $\chi$. Here, the response is that of a non-interacting
system, $\Pi(\ve{q},\omega)$, to an external potential which includes
an effective potential due to the perturbed density, i.e., one
replaces $V^{ext}(\ve{q},\omega)$ in \eqref{eq:linear_response} with
$V^{ext}(\ve{q},\omega) + v(\ve{q}) \delta n(\ve{q},\omega)$, giving
$\chi_{RPA}^{-1}(\ve{q},\omega) = \Pi^{-1}(\ve{q},\omega)-v(\ve{q})$.
However, as discussed below, RPA is never accurate for dipolar
interactions.

We account for correlations beyond RPA by including a local field
factor $G(\ve{q})$ in the response function~\cite{vignale_book}:
\begin{equation}
  \chi(\ve{q},\omega) = \frac{\Pi(\ve{q},\omega)}{1-v(\ve{q})
    \left[1-G(\ve{q})\right] \Pi(\ve{q},\omega)}\; .
\label{eq:chire}
\end{equation}
%   
%More generally, $G$ will also have a frequency dependence, but the
%static approximation already gives us considerable insight into the
%problem.
Physically, $G(\ve{q})$ corresponds to the corrections to the RPA
effective potential that stem from correlations between fermions. For
example, at short distances (large $q$) the interactions will be
suppressed by Pauli exclusion, thus giving $G=1$. 
These exchange correlations, which are crucial in a gas of identical fermions, 
are clearly neglected by RPA. 
In addition, we can also extract the
behavior of $G$ in the opposite limit $q\to 0$ using the
compressibility sum rule~\cite{vignale_book,NozPine}, which relates
$\chi^{-1}(\ve{q}\to 0,0)$ to the inverse compressibility $\kappa^{-1}
= n^2 \partial^2 (n\varepsilon)/\partial n^2$, where $\varepsilon$ is
the ground state energy per particle.
For Coulomb interactions in electron systems, where
  $v(\ve{q})\propto 1/q$, the Hartree-Fock calculation for
$\varepsilon$ gives us $G(\ve{q}) \simeq 10q/(3\pi k_F)$ as $q \to 0$,
thus confirming that $G \to 0$: RPA is therefore a reasonable
  approximation for long wavelengths~\cite{vignale_book}.
%Indeed, we expect $G(\ve{q}) \propto q$ generally for $\ve{q} \to 0$
%in order to cancel the $1/q$ divergence in the Coulomb potential.
This is not however true in the case of dipolar interactions: If we
perform the same procedure, where we take $\theta = 0$ in
Eq.~\eqref{eq:dipolar_int} for simplicity, then we find that $G(0) = 1
- 32\hbar^2 U/(3mV_0)$ in the limit $U \ll 1$, where the Hartree-Fock
result \eqref{eq:HF} is valid. Thus we see that $\chi_{RPA}$ is never
recovered in this case, even in the weak-coupling limit. In sum, the
RPA for 2D dipolar Fermi gases fails at both short and long
wavelengths~\footnote{Note that, already in Ref.~\cite{zinner2011} it
  was recognised that exchange correlations beyond RPA must be
  incorporated in the dipolar gas. However, they consider a
  phenomenological expression for the local field factor with a form
  similar to that in Coulomb systems.}.

We instead determine $G(\ve{q})$ using the STLS scheme, which provides
an ingenious way in which to feed back the correlations in
$\chi(\ve{q},\omega)$ into $G(\ve{q})$.  STLS uses a classical analogy
for the system's response to obtain~\cite{STLSpaper}
\begin{align}
  G(\ve{q}) = -\frac{1}{n} \int \frac{d\ve{k}}{(2\pi)^2}
  \frac{\ve{q}\cdot\ve{k}}{q^2} \frac{v(\ve{k})}{v(\ve{q})} \lbc
  S(\ve{q}-\ve{k}) - 1 \rbc \; .
\label{eq:local_field}
\end{align}
Note that the RPA case of $G=0$ implies that $S(\ve{q})=1$ here, which
in turn corresponds to setting $g(\ve{r}) = 1$, i.e. neglecting any
correlations in the STLS classical analogy.
Combining Eq.~\eqref{eq:local_field} with Eq.~\eqref{eq:struct_fact}
gives us a set of self-consistent equations for $G(\ve{q})$ that can
be solved iteratively. If we start by inserting $S_0(\ve{q})$ into
Eq.~\eqref{eq:local_field} (which is equivalent to setting
$G^{(0)}(\ve{q})=1$ at the beginning of the iteration), then
$G^{(1)}(\ve{q})$ incorporates exchange correlations only.
In particular, if we have purely contact interactions $v(\ve{q}) =
V_0$, then Eq.~\eqref{eq:local_field} returns $G^{(1)}(\ve{q}) = 1-
g(0) = 1$. Thus, we see that STLS correctly gives us a non-interacting
response for a gas of identical fermions with contact interactions.

For the dipolar interaction~\eqref{eq:dipolar_int}, one can show that
$G(\ve{q})$ calculated from Eq.~\eqref{eq:local_field} will also
render $\chi(\ve{q},\omega)$ independent of $V_0$ provided
$g(0)=0$. However, similarly to what happens in the electronic Coulomb
case~\cite{vignale_book}, the STLS scheme does not guarantee that
$g(0) = 0$ for the converged solution and so we
%often
obtain an unphysical dependence on $V_0$. In addition, we find that
the density instabilities determined using this procedure are
sensitive to the cut-off $\Lambda$ at large $q$ even though we have
$q_c \leq 2 k_F$.

To address these issues and better model the dipolar gas, we improve
the STLS scheme by imposing, at each iteration step, the constraint
$g(0) = 0$ and the fact that $\chi(\ve{q},\omega)$, and thus
$S(\ve{q})$, will be dominated by Pauli exclusion for $q\gg 2 k_F$.
Similarly to Ref.~\cite{yoshizawa2009}, we achieve this by adding a
corrective function $\delta S(\ve{q})$ to the $S(\ve{q})$ defined by
Eq.~\eqref{eq:struct_fact} and then using $S+\delta S$ to determine
$G(\ve{q})$. In particular, we use the ansatz
\begin{equation*}
  \delta S(\ve{q}) = \lba S_0(\ve{q}) - S(\ve{q}) + A e^{-q^2/(2k_F)^2}
  \rba \lba 1 - e^{-q^2/(2k_F)^2}\rba
\end{equation*}
to interpolate between the STLS result for $q < 2k_F$ and the
non-interacting one $S_0$ for $q\gg 2k_F$, where exchange correlations
dominate.
%We choose $q_0 = 2 k_F$ as the relevant momentum scale, 
%
The constant $A$ adjusts the behavior near $q = 2k_F$ and is chosen at
each iteration step so that $g(0) = 0$.  Note that the correction
around $q\simeq 2k_F$ is generally small. Our improved STLS procedure
thus renders $\chi(\ve{q},\omega)$ insensitive to both $V_0$ and
cut-off $\Lambda \gg 2 k_F$, as required.

We have confirmed that our converged solution for $U\ll 1$ agrees with
the weak-coupling Hartree-Fock result. In this limit, the Hartree-Fock
approximation for the dipolar gas gives us a ground-state energy per
particle:
\begin{align}
  \varepsilon_{HF} = \frac{\hbar^2 k_F^2}{m} \lbc \frac{1}{4} +
  \frac{16}{45\pi} U(3\cos^2\theta - 1) \rbc \; .
\label{eq:HF}
\end{align}
Here, we only consider up to first-order in $U$ for the energy density
$\varepsilon_{HF}$, and thus we have neglected the higher order terms
due to Fermi surface deformations induced when $\theta \neq
0$~\cite{yamaguchi2010}.
We compare this expression with the ground state energy density
extracted from our STLS solution for $\chi(\ve{q},\omega)$ using the
following relation for the interaction energy per particle:
\begin{align}
  \varepsilon_{int} = \frac{n}{2} v(0) + \frac{1}{2} \int
  \frac{d\ve{q}}{(2\pi)^2} v(\ve{q}) \lbc S(\ve{q}) -1 \rbc
\label{eq:int_erg}
\end{align}
and then employing the Hellman-Feynman
theorem~\cite{NozPine,vignale_book}. By doing this, we find that the
ground state energy density obtained via the STLS calculation recovers
the Hartree-Fock result~\eqref{eq:HF} when $U \ll 1.$ Equivalently,
we recover 
%the Hartree-Fock energy density
$\varepsilon_{HF}$ if we impose $S(\ve{q}) = S_0(\ve{q})$ in
Eq.~\eqref{eq:int_erg}.

\begin{figure}
\centering
\includegraphics[width=0.8\linewidth,angle=0]{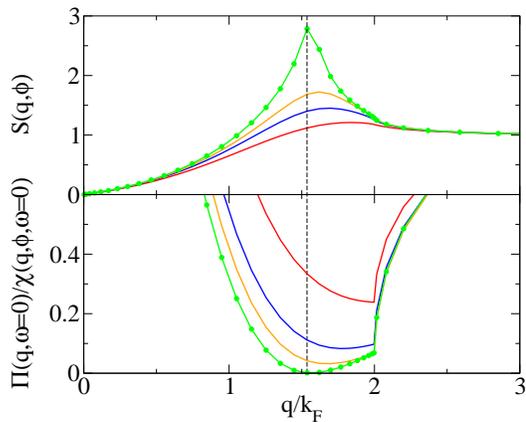}
\caption{(Color online) Behaviour of the static structure factor
  $S(q,\phi)$ and the rescaled inverse density-density response
  function, $\Pi(q,\omega=0)/\chi(q,\phi;\omega=0)$, as the stripe
  instability at $\phi=\pi/2$ is approached for $\theta=0.1$. The
  vertical dashed line marks the position of the unstable wave vector
  $q_c \simeq 1.54 k_F$.}
\label{fig:stripe_ins}
\end{figure}

Using our procedure, we analyse the density instabilities of the
converged solutions for $\chi(\ve{q},\omega)$. For tilted dipoles
($\theta \neq 0$), $\chi(\ve{q},0)$ is most unstable towards forming a
density wave along $\phi=\pi/2$, as shown in the
Fig.~\ref{fig:phase_diag} inset.
Referring to the phase diagram in Fig.~\ref{fig:phase_diag}, we find
that this stripe phase exists for sufficiently large $U$ when $\theta
\lesssim \pi/4$. RPA also predicts a stripe transition for $1/U =
2\cos^2\theta$ once one sets $V_0=0$ in Eq.~\eqref{eq:dipolar_int}
(cf.\ Refs.~\cite{yamaguchi2010,DasSarma2010}). However, we see that
correlations shift the transition to a much higher $U$ compared to the
RPA result, thus giving $p$-wave superfluidity~\cite{bruun2008} a
sizeable region of existence around $\theta = \pi/4$.  Moreover, we
find that $q_c < 2 k_F$ rather than $q_c = 2k_F$ as expected from RPA.
Figure~\ref{fig:stripe_ins} shows how $\chi(\ve{q},0)^{-1}$ tends
toward zero (i.e.\ how $\chi(\ve{q},0)$ diverges) as we approach the
stripe transition at fixed $\theta$. The divergence in
$\chi(\ve{q},0)$ leads to a singularity in Eq.~\eqref{eq:struct_fact},
thus yielding a corresponding peak in the structure factor $S(q,\phi)$
at $q=q_c$.  Once $\chi(\ve{q},0)^{-1}$ hits zero at the stripe
transition, we find that we no longer obtain convergence of the
self-consistent equations \eqref{eq:struct_fact} and
\eqref{eq:local_field} when we increase $U$ further.

For the isotropic case ($\theta = 0$), one might expect the
inhomogeneous phase to maximise its rotational symmetry by forming a
triangular lattice.
%{\tt (cite Brazovskii?)} 
However, we instead find that the system spontaneously breaks
rotational symmetry to form a stripe phase. We see this by setting
$G^{(0)}(\ve{q})$ to a converged solution for small $\theta$ and $U$,
and then examining whether or not the iteration procedure for $\theta
= 0$ amplifies or suppresses the spread in $\phi$. From the final
converged solutions, we find that $\chi(\ve{q},0)$ exhibits a large
spread in $\phi$ at $U=6$, before eventually diverging for a specific
$\phi$ at $U\simeq 6.03$. Here, the direction of the stripe is simply
determined by the original $\phi$ dependence of
$G^{(0)}(\ve{q})$. Thus, we see that the system is unstable towards
breaking rotational symmetry and spontaneously forming a stripe
phase~\footnote{Note that this implicitly assumes that the stripe
  transition is second order, but it could instead be preempted by a
  first order transition --- one needs to consider higher order terms
  in $\delta n$ in the free energy to ascertain this.}.  There is also
the possibility that the system first forms a nematic phase, similar
to that discussed in Ref.~\cite{fregoso2009}, before forming a stripe
phase.

If we neglect any dependence on $\phi$ and consider $G$ and $S$ to be
functions of $q$ only, then we never see a transition to an
inhomogeneous phase.  However, we do see evidence of strong
correlations in the pair correlation function
(Fig.~\ref{fig:pair_corr}) and a peak in the structure factor that
suggests an imminent transition.  Such a behavior is consistent with a
first-order transition to a Wigner crystal phase. If this transition
is similar to that of dipolar bosons, then it is expected to occur at
$U\simeq 60$~\cite{astrakharchik_07,buchler2007,mora2007}.

As $\theta$ increases towards $\pi/2$ in Fig.~\ref{fig:phase_diag}, we
instead find that the system can be unstable towards collapse, where
$q_c = 0$ at the instability. In this case, $\chi(\ve{q},0)$ is most
unstable for $\phi=0$, which implies that the system executes an
anisotropic collapse in the direction of the dipole tilt, a physically
reasonable scenario.
Contrast this with RPA, where one only ever has $q_c > 0$.
Sun et al.~\cite{DasSarma2010} use a perturbative expansion of
$\chi(\ve{q},0)$ around $\ve{q}=0$ to argue that one never has
instabilities with $q_c = 0$ in the 2D dipolar gas. However, their
argument rests on the assumption that $\chi(\ve{q},0)$ is analytic at
$\ve{q}=0$, as is the case with RPA, while we find that our
$\chi(\ve{q},0)$ depends on $\phi$ at $\ve{q}=0$ and is thus
non-analytic.  This non-analytic behavior at $\ve{q}=0$ may, at first
sight, appear surprising for a Fermi liquid, but it merely corresponds
to an anisotropic compressibility, which is physically reasonable for
anisotropic interactions. Moreover, it is consistent with the collapse
predicted from Hartree-Fock
calculations~\cite{bruun2008,yamaguchi2010,sieberer2011,babadi2011}, as
depicted in Fig.~\ref{fig:phase_diag}. We recover the Hartree-Fock
calculations for the collapse by using in Eq.~\eqref{eq:chire} the
exchange-only field factor $G^{(1)} (\ve{q})$ previously discussed.
The Hartree-Fock approximation should be accurate for the collapse
instability since the instability simply corresponds to the point at
which the attractive interaction exceeds the effective repulsive
interaction derived from Pauli exclusion.  However, we do not expect
such an approximation to be accurate for the stripe instability since
we require correlations beyond the exchange ones in this case.
Indeed, the conserving Hartree-Fock calculation employed in
Refs.~\cite{sieberer2011,babadi2011} is expected to underestimate the
interaction at which the stripe transition occurs, as stressed in
Ref.~\cite{babadi2011}. This explains the quantitative disagreement
between our phase boundary and that of
Refs.~\cite{sieberer2011,babadi2011}.

%\paragraph{Discussion ---}
Despite the apparent success of our improved STLS scheme for the
dipolar gas, there are still some inconsistencies that it shares with
the original STLS scheme for electron systems.  Specifically, the pair
correlation function can become slightly negative at short distances
(Fig.~\ref{fig:pair_corr}) and the compressibility sum rule is
systematically violated
for a range of interaction strengths $U \lesssim 3$. However, our
scheme is a substantial improvement over RPA and we expect it to
provide a basis upon which to investigate correlations in other
dipolar Fermi systems such as multilayers.

Our predicted stripe phases should be experimentally realizable with
polar molecules, where the density modulations could be probed using
Bragg scattering.
The typical density of polar molecules in a 2D layer is $10^8$
cm$^{-2}$, which gives a maximum of $U\simeq 0.3$ for KRb molecules
with dipole moment $D \sim 0.2$ Debye as in the
experiment~\cite{miranda2011}.  Thus, to access the stripe phase with
current experiments, one needs to enhance $U$ by, e.g., increasing the
effective mass using an in-plane optical lattice. Alternatively, one
could use LiCs molecules which have a dipole moment of up to 5.5
Debye~\cite{Carr2009}, thus allowing one to explore the stripe
transition for the whole range of $\theta$.

%%%%%%%%%%%%%%%%%%%%%%%%%%%%%%%%%%%%%%%%%%%%%%%%%
\acknowledgments We are grateful to P. Littlewood, M. Polini and
N. Zinner for useful discussions. MMP acknowledges support from the
EPSRC under Grant No.\ EP/H00369X/1. FMM acknowledges financial
support from the programs Ram\'on y Cajal, Polatom (ESF), and CAM
(S-2009/ESP-1503).
% 
%%%%%%%%%%%%%%%%%%%%%%%%%%%%%%%%%%%%%%%%%%%%%%%%%

%\bibliography{dipoleRefs}

\end{document}